\def\ifarxiv{\iftrue}     

\ifarxiv
\documentclass[12pt,a4paper]{article}

\usepackage{amsmath,amssymb}
\usepackage[nosort]{cite}
\usepackage[hyperref,parsep]{collect}

\makeatletter
\let\old@makecaption=\@makecaption
\def\@makecaption{\small\old@makecaption}
\makeatother

\else

\documentclass[twoside,reqno]{qts-proc}
\usepackage{cite}
\usepackage{amssymb,amsmath}
\usepackage{times}
\fi

\makeatletter
\let\old@startsection=\@startsection
\renewcommand{\@startsection}[6]{\old@startsection{#1}{#2}{#3}{#4}{#5}{#6\mathversion{bold}}}
\makeatother

\let\oldPhi=\Phi
\let\oldPsi=\Psi
\let\oldGamma=\Gamma
\let\oldDelta=\Delta
\let\oldSigma=\Sigma
\let\oldTheta=\Theta
\let\oldPi=\Pi
\renewcommand{\Phi}{\mathnormal{\oldPhi}}
\renewcommand{\Psi}{\mathnormal{\oldPsi}}
\renewcommand{\Gamma}{\mathnormal{\oldGamma}}
\renewcommand{\Sigma}{\mathnormal{\oldSigma}}
\renewcommand{\Delta}{\mathnormal{\oldDelta}}
\renewcommand{\Theta}{\mathnormal{\oldTheta}}
\renewcommand{\Pi}{\mathnormal{\oldPi}}

\newcommand{\ham}{\mathcal{H}}
\newcommand{\smat}{\mathcal{S}}
\newcommand{\perm}{\mathcal{P}}
\newcommand{\ident}{\mathcal{I}}
\newcommand{\rmat}{\mathcal{R}}

\newcommand{\gen}[1]{\mathfrak{#1}}

\newcommand{\superN}{\mathcal{N}}
\newcommand{\fldZ}{\mathcal{Z}}

\newcommand{\cdott}{\mathord{\cdot}}

\ifx\genfrac\sdflkaj
\newcommand{\atopfrac}[2]{{{#1}\above0pt{#2}}}
\else
\newcommand{\atopfrac}[2]{\genfrac{}{}{0pt}{}{#1}{#2}}
\fi
\newcommand{\sfrac}[2]{{\textstyle\frac{#1}{#2}}}
\newcommand{\half}{\sfrac{1}{2}}
\newcommand{\ihalf}{\sfrac{i}{2}}
\newcommand{\quarter}{\sfrac{1}{4}}

\newcommand{\rep}[1]{{\mathbf{#1}}}

\newcommand{\alg}[1]{\mathfrak{#1}}

\newcommand{\lrbrk}[1]{\left(#1\right)}
\newcommand{\bigbrk}[1]{\bigl(#1\bigr)}

\newcommand{\comm}[2]{[#1,#2]}

\newcommand{\acomm}[2]{\{#1,#2\}}

\newcommand{\state}[1]{\mathopen{|}#1\mathclose{\rangle}}

\newcommand{\nln}{\nonumber\\}
\newcommand{\nl}[1][0pt]{\nonumber\\[#1]&\hspace{-4\arraycolsep}&\mathord{}}

\newcommand{\earel}[1]{\mathrel{}&\hspace{-2\arraycolsep}#1\hspace{-2\arraycolsep}&\mathrel{}}
\newcommand{\eq}{\earel{=}}

\def\[{\begin{equation}}
\def\]{\end{equation}}
\def\<{\begin{eqnarray}}
\def\>{\end{eqnarray}}

\newcounter{enumlistcnt}
\renewcommand{\theenumlistcnt}{\protect\emph{\roman{enumlistcnt}}}

\newcommand{\MMM}[2]{{\arraycolsep0pt\begin{array}[b]{c}\makebox[0cm]{$\atopfrac{#2}{\downarrow}$}\\#1\end{array}}}

\makeatletter
\def\mr@ignsp#1 {\ifx\:#1\@empty\else #1\expandafter\mr@ignsp\fi}%
\newcommand{\multiref}[1]{\begingroup
\xdef\mr@no@sparg{\expandafter\mr@ignsp#1 \: }%
\def\mr@comma{}%
\@for\mr@refs:=\mr@no@sparg\do{\mr@comma\def\mr@comma{,}\ref{\mr@refs}}%
\endgroup}
\makeatother

\newcommand{\hypref}[2]{\ifx\href\asklfhas #2\else\href{#1}{#2}\fi}

\renewcommand{\eqref}[1]{(\multiref{#1})}

\ifx\href\asklfhas\newcommand{\href}[2]{#2}\fi
\newcommand{\arxivno}[1]{\href{http://arxiv.org/abs/#1}{#1}}

\begin{document}

\ifarxiv

\begin{flushright}\footnotesize
\texttt{\arxivno{hep-th/0511013}}\\
\texttt{PUTP-2180}
\end{flushright}
\vspace{0cm}

\begin{center}
{\Large\textbf{\mathversion{bold}
An $\alg{su}(1|1)$-Invariant S-Matrix\\
with Dynamic Representations
}\par}
\vspace{1cm}

\textsc{Niklas Beisert}
\vspace{5mm}

\textit{Joseph Henry Laboratories\\
Princeton University\\
Princeton, NJ 08544, USA}\vspace{3mm}
\vspace{3mm}

\texttt{nbeisert@princeton.edu}\par\vspace{1cm}

\textbf{Abstract}\vspace{7mm}

\begin{minipage}{12.7cm}
The spin chains originating from large-$N$ conformal gauge theories 
are of a special kind:
The Hamiltonian is \emph{not invariant}
under the symmetry algebra, it is rather \emph{a part} of it.
This leads to interesting properties within the asymptotic Bethe ansatz. 
Here we study an S-matrix with $\alg{u}(1|1)$ symmetry
which arises in a long-range spin chain 
with fundamental spins of $\alg{su}(2|1)$.
\end{minipage}

\end{center}
\vspace{1cm}
\hrule height 0.75pt
\vspace{1cm}

\else

\sloppy \raggedbottom
\setcounter{page}{1}

\newpage
\setcounter{figure}{0}
\setcounter{equation}{0}
\setcounter{footnote}{0}
\setcounter{table}{0}
\setcounter{section}{0}

\title{\mathversion{bold}An $\alg{su}(1|1)$-Invariant S-Matrix\\ 
with Dynamic Representations}

\runningheads{N.~Beisert}
{An su(1/1)-Invariant S-Matrix 
with Dynamic Representations}

\begin{start}

\author{N.~Beisert}{1}

\address{Joseph Henry Laboratories\\
Princeton University\\
Princeton, NJ 08544, USA}{1}

\begin{Abstract}
The spin chains originating from large-$N$ conformal gauge theories 
are of a special kind:
The Hamiltonian is \emph{not invariant}
under the symmetry algebra, it is rather \emph{a part} of it.
This leads to interesting properties within the asymptotic Bethe ansatz. 
Here we study an S-matrix with $\alg{u}(1|1)$ symmetry
which arises in a long-range spin chain 
with fundamental spins of $\alg{su}(2|1)$.
\end{Abstract}
\end{start}

\fi

\section{Introduction}

Field theories with extended supersymmetries have turned 
out to be a reliable source for unexpected further symmetries.
The most famous examples are probably the maximally extended 
supergravities \cite{Cremmer:1978km} 
with exceptional groups as hidden symmetries
\cite{Cremmer:1978ds,Cremmer:1979up,deWit:2000wu}.
Similarly, maximally supersymmetric $\superN=4$ gauge
theory in four dimensions 
\cite{Gliozzi:1977qd,Brink:1977bc}
has been known for a long time
to be one of the few finite quantum field theories \cite{Sohnius:1981sn,Howe:1984sr,Brink:1983pd}. 
Due to masslessness, the quantum theory is superconformal
and has the global symmetry algebra $\alg{psu}(2,2|4)$.
In recent years it has become clear that there is even more
symmetry when restricting to the large-$N$ limit:
\emph{Integrability}. In the study of the AdS/CFT correspondence 
\cite{Maldacena:1998re,Gubser:1998bc,Witten:1998qj},
Minahan and Zarembo found an integrable
structure \cite{Minahan:2002ve} which was subsequently extended to
higher perturbative orders \cite{Beisert:2003tq,Beisert:2003ys} 
and all local operators \cite{Beisert:2003yb}.%
\footnote{Lipatov's remarks on integrability in $\superN=4$ SYM \cite{Lipatov:1997vu,Lipatov:1998as}
were well ahead of these developments.}
Due to the apparent integrability, 
planar anomalous dimensions can now be computed by means of Bethe ans\"atze
\cite{Minahan:2002ve,Beisert:2003tq,Serban:2004jf,Beisert:2004hm,Staudacher:2004tk,Beisert:2005fw}.
The latter have proved to be extremely useful for studies of the AdS/CFT
correspondence \cite{Beisert:2003xu,Beisert:2003ea}.
For reviews of the subject of integrability in $\superN=4$ SYM, 
please refer to \cite{Beisert:2004ry,Beisert:2004yq,Zarembo:2004hp,Plefka:2005bk}.

In \cite{Beisert:2005fw} an all-loop solution was proposed 
for spectrum of the $\alg{su}(1|2)$ sector of $\superN=4$ SYM. 
This solution involved an interesting S-matrix 
obeying the Yang-Baxter relation. 
The S-matrix was `extrapolated' from 
its first few perturbative orders which were derived 
from the underlying Hamiltonian. 
It has a remarkably simple form, but its nature mostly remained obscure.
It is one of the purposes of this note to clarify 
the origin of the S-matrix from a representation 
theory point of view.
For instance, in most known cases the S-matrix 
is uniquely determined by the symmetries it obeys 
and the representations of the scattering objects. 
This will turn out to be the case here as well. 

\section{The $\alg{su}(1|2)$ Sector of $\superN=4$ SYM}

The $\alg{su}(1|2)$ sector of $\superN=4$ SYM 
is equivalent to a long-range spin chain 
where each spin takes one out of three orientations:
The bosonic orientation $\fldZ$ is considered to
be the vacuum while the other two orientations
$\phi$ and $\psi$, which are bosonic and
fermionic, respectively, are considered 
to be the excitations of the vacuum.
The Hamiltonian of this spin chain is of the 
long-range type as introduced in \cite{Beisert:2003tq}.
In other words, the Hamiltonian
\[
\ham(\lambda)=\ham_0+\lambda \ham_1+\lambda^2 \ham_2+\ldots
\]
is a perturbative deformation 
of a nearest-neighbour spin interaction $\ham_0$ with 
the higher orders $\ham_r$ 
in the small coupling $\lambda$ 
acting among $r+2$ adjacent spins.
For the given sector the Hamiltonian was
computed up to second order in \cite{Beisert:2003ys}.

\section{The Asymptotic Bethe Ansatz}

To construct the integrable structure of such a long-range system
remains a difficult task. The R-matrix suits this purpose
well only for common nearest-neighbour spin chains;
it is not yet clear if and how it might be applied here.
Nevertheless, these difficulties can be overcome 
by considering asymptotic states and 
the S-matrix of the excitations \cite{Sutherland:1978aa,Staudacher:2004tk}.
In an asymptotic state all excitations 
are sufficiently separated along the spin chain, e.g.
\[\label{eq:asymptotic}
\state{\phi_1 \ldots \psi_K}
=\sum_{\makebox[0pt]{$\scriptstyle a_1\ll\ldots\ll a_K$}}e^{ia_1p_1+\ldots+ia_K p_K}\,
\state{\ldots\fldZ\fldZ\ldots \MMM{\phi}{a_1}\ldots \MMM{\ldots}{\ldots}\ldots 
\MMM{\psi}{a_K}\ldots \fldZ\fldZ\ldots}.
\]
Here, the indices $k=1,\ldots,K$ refer to the momenta $p_k$ of the excitation.
The crucial insight is that the Hamiltonian is 
\emph{homogeneous} and \emph{local}. These asymptotic states are
therefore asymptotically eigenstates of the Hamiltonian.
Homogeneity leads to the plane wave factors in \eqref{eq:asymptotic} 
and locality makes the propagation of the individual excitations
independent of each other. 
The only violation of the eigenstate equation for 
the state in \eqref{eq:asymptotic} comes from the vicinity 
of the boundaries of the asymptotic region; 
when two excitations come too close they will interact non-trivially. 

In order to construct exact eigenstates, the various asymptotic
regions have to be stitched up in a suitable way. 
For an integrable system, this can be achieved by a pairwise S-matrix.
For instance, to combine the two asymptotic regions where
the first two excitations are in either ordering, one would use
\[
(1+\smat_{12})\,\state{\phi_{1}\phi_2\ldots}
=
\state{\phi_1\phi_2\ldots}
+S_{12}\state{\phi_2\phi_1\ldots}.
\]
Here $S_{12}$ represents some phase factor due to the application of the S-matrix. 
A generic asymptotic eigenstate%
\footnote{An asymptotic eigenstate is defined to
be a state which can be completed by non-asymptotic contributions 
(nearby excitations) to obtain an exact eigenstate.}
can be constructed as follows
\[
\state{\Psi}=\sum_{\pi\in S_K} \smat_\pi\, \state{\phi_1\ldots\psi_K},
\]
where $\smat_\pi$ is a product of nearest-neighbour S-matrices 
which permute according to the permutation $\pi$.

\section{The $\alg{su}(1|1)$ Asymptotic Symmetry Algebra}

The Hamiltonian is part of the symmetry algebra $\alg{su}(2|1)$
which acts on the spin chain. The residual algebra which leaves 
the number of excitations invariant is $\alg{u}(1|1)$. 
It consists of the two supercharges $\gen{Q},\gen{S}$,
the outer automorphism $\gen{B}$ and the central charge $\gen{C}$. 
The central charge contains the Hamiltonian
\[
\gen{C}(\lambda)=\gen{C}_0+\lambda\, \ham(\lambda).
\]
The non-trivial commutators of the $\alg{u}(1|1)$ algebra are given by 
\[
\acomm{\gen{Q}}{\gen{S}}=\gen{C},\quad
\comm{\gen{B}}{\gen{Q}}=-2\gen{Q},\quad
\comm{\gen{B}}{\gen{S}}=+2\gen{S},\quad
\]
The algebra also has an invariant quadratic 
Casimir operator $\gen{J}^2$, it reads
\[
\gen{J}^2=2\comm{\gen{Q}}{\gen{S}}+\acomm{\gen{B}}{\gen{C}}.
\]

We can now construct a representation on a single excitation. 
The most general solution of the algebra relations is given by
\[\label{eq:extrans}
\begin{array}[b]{rclcrcl}
\gen{B}\,\state{\phi}\eq (b+1)\state{\phi},&&
\gen{B}\,\state{\psi}\eq (b-1)\state{\psi},\\[3pt]
\gen{Q}\,\state{\phi}\eq q\, \state{\psi},&&
\gen{Q}\,\state{\psi}\eq 0,\\[3pt]
\gen{S}\,\state{\phi}\eq 0,&&
\gen{S}\,\state{\psi}\eq c/q\, \state{\phi},\\[3pt]
\gen{C}\,\state{\phi}\eq c\,\state{\phi},&&
\gen{C}\,\state{\psi}\eq c\,\state{\psi},\\[3pt]
\gen{J}^2\,\state{\phi}\eq 2bc\,\state{\phi},&&
\gen{J}^2\,\state{\psi}\eq 2bc\,\state{\phi}.
\end{array}
\]
The coefficient $q$ is an unphysical quantity
which reflects the difference of normalizations between
bosonic and fermionic excitations.
Note that the representation agrees with the 
expansion of generators derived in \cite{Beisert:2003ys} 
when applied to single excitations restricted to the $\alg{su}(1|2)$ sector. 
There, $c$ represents the energy and $q$ contains 
some of the unphysical constants related to similarity transformations.

Let us denote the above representation \eqref{eq:extrans} 
with central charge $c$ and hypercharge $b$ 
by $\rep{(1|1)}_{c,b}$. 
The representation on asymptotic states with $K$ excitations
is the tensor product 
$\rep{(1|1)}_{c_1,b_1}\otimes \rep{(1|1)}_{c_2,b_2}\otimes\ldots\otimes \rep{(1|1)}_{c_K,b_K}$.

\section{The Invariant S-Matrix}

The S-matrix is an invariant operator acting on 
two modules by interchanging them
\[
\smat_{12}:\rep{(1|1)}_{c_1}\otimes \rep{(1|1)}_{c_2}
\to\rep{(1|1)}_{c_2}\otimes \rep{(1|1)}_{c_1}.
\]
We shall write it as a product of an operator $\rmat_{12}$ (R-matrix)
and the (graded) permutation $\perm_{12}$
\[
\smat_{12}=\perm_{12}\,\rmat_{12}(a_1,a_2),\qquad
\rmat_{12}:\rep{(1|1)}_{c_1}\otimes \rep{(1|1)}_{c_2}
\to\rep{(1|1)}_{c_1}\otimes \rep{(1|1)}_{c_2}.
\]
The R-matrix depends on two spectral parameters
$a_k=a(p_k)$, which are themselves function of the 
particle momenta.
The permutation is clearly $\alg{u}(1|1)$ invariant 
and the same must therefore be true for the R-matrix.
The latter can therefore be written as
a sum over projectors to irreducible components.
The tensor product in question decomposes into two similar
irreducible modules $\rep{(1|1)}_{c_1+c_2,b_1+b_2+1}$
and $\rep{(1|1)}_{c_1+c_2,b_1+b_2-1}$.
These two modules can be distinguished by the
quadratic Casimir on the tensor product
\[
\gen{J}_{12}^2=\gen{J}_1^2+2\gen{J}_1\cdott\gen{J}_2+\gen{J}_2^2=
2b_1c_1+2b_2c_2+
2\gen{B}_1\gen{C}_2+2\gen{C}_1\gen{B}_2+4\gen{Q}_1\gen{S}_2-4\gen{S}_1\gen{Q}_2.
\]
Because there are only two irreps in the tensor product, 
all invariant operators 
can now be written as a linear combination of the identity operator
$\ident_{12}$ and the quadratic Casimir $\gen{J}_{12}^2$.
In particular this applies to the square of $\gen{J}_{12}^2$
\[
(\gen{J}_{12}^2)^2=4(b_1+b_2)(c_1+c_2)\,\gen{J}_{12}^2
-4(b_1+b_2+1)(b_1+b_2-1)(c_1+c_2)^2\,\ident_{12}.
\]
%

The same applies to the R-matrix which we can write as 
\[
\rmat_{12}(a_1,a_2)=R_{12,1}(a_1,a_2) \,\ident_{12}+R_{12,2}(a_1,a_2)\, \gen{J}_{12}^2
\]
with some coefficients $R_{12,1},R_{12,2}$ to be determined.
It should obey the unitarity and Yang-Baxter relations 
\[
\rmat_{12}\rmat_{21}=\ident_{12},\qquad
\rmat_{12}\rmat_{13}\rmat_{23}=\rmat_{23}\rmat_{13}\rmat_{12}.
\]
The standard solution for the coefficients in the R-matrix is 
\<
R_{12,1}(a_1,a_2)\eq
\frac{a_2-a_1-\sfrac{i}{2}(b_1+b_2)(c_1+c_2)}{a_2-a_1-\sfrac{i}{2}(c_1+c_2)}\,R_{12,0}(a_1,a_2),\nln
R_{12,2}(a_1,a_2)\eq
\frac{\sfrac{i}{4}}{a_2-a_1-\sfrac{i}{2}(c_1+c_2)}\,R_{12,0}(a_1,a_2)
\>
with some undetermined phase $R_{12,0}(a_1,a_2)$
obeying 
\[
R_{12,0}(a_1,a_2)R_{12,0}(a_2,a_1)=1.
\]
Let us for convenience drop this overall phase, $R_{12,0}=1$.
We apply the R-matrix to a two-particle state and find
\<
\rmat_{12}(a_1,a_2)\,\state{\phi_1\phi_2}\eq 
\frac{a_2-a_1+\sfrac{i}{2}(c_1+c_2)}{a_2-a_1-\sfrac{i}{2}(c_1+c_2)}\,\state{\phi_1\phi_2},
\nln
\rmat_{12}(a_1,a_2)\,\state{\phi_1\psi_2}\eq 
\frac{a_2-a_1+\sfrac{i}{2}(c_2-c_1)}{a_2-a_1-\sfrac{i}{2}(c_1+c_2)}\,\state{\phi_1\psi_2}
\ifarxiv\else\nl\fi
+\frac{ic_2}{a_2-a_1-\sfrac{i}{2}(c_1+c_2)}\,\frac{q_1}{q_2}\,\state{\psi_1\phi_2},
\nln
\rmat_{12}(a_1,a_2)\,\state{\psi_1\phi_2}
\eq
\frac{a_2-a_1+\sfrac{i}{2}(c_1-c_2)}{a_2-a_1-\sfrac{i}{2}(c_1+c_2)}\,\state{\psi_1\phi_2}
\ifarxiv\else\nl\fi
+\frac{ic_1}{a_2-a_1-\sfrac{i}{2}(c_1+c_2)}\,\frac{q_2}{q_1}\,\state{\phi_1\psi_2},
\nln
\rmat_{12}(a_1,a_2)\,\state{\psi_1\psi_2}\eq 
\frac{a_2-a_1-\sfrac{i}{2}(c_1+c_2)}{a_2-a_1-\sfrac{i}{2}(c_1+c_2)}\,\state{\psi_1\psi_2}.
\>
More general results for quantum deformed symmetry algebras can 
be found in \cite{Kulish:1989sv,Reshetikhin:1990ep,Dabrowski:1991nr,Hinrichsen:1991nj,Reshetikhin:1992aa,Bracken:1994hz,Delius:1994yn,Zhang:1998bn}.
Note that above R-matrix coincides with the results in \cite{Bracken:1994hz,Delius:1994yn,Zhang:1998bn}
when setting $\alpha,\beta=c_1,c_2$, $q=\exp(\kappa)$, $x=\exp(2i\kappa(a_2-a_1))$
and sending the deformation parameter $\kappa\to 0$.

\section{The S-Matrix for the $\alg{su}(1|2)$ Sector}

As the next step, we apply the above results to 
asymptotic states of the spin chain for $\superN=4$ SYM.
There, the spectral parameter $a$ of a particle
is given by a function of the momentum,
$a=a(p)$. Furthermore, the central charge is interpreted 
as the energy of a state as $c=1+\lambda e$
which itself is given through the dispersion relation $e=e(p)$,
i.e.~we have $c=c(p)$. We can now perform a change of parameters
as follows
\[a=\half(x^++x^-),\qquad c=-i(x^+-x^-),\]
where now $x^\pm$ are given as some functions of
the momentum, $x^\pm=x^\pm(p)$. Substituting this in 
the S-matrix, we obtain 
\<\label{eq:su11s}
\smat_{12}\,\state{\phi_1\phi_2}\eq 
\frac{x^+_2-x^-_1}{x^-_2-x^+_1}\,\state{\phi_2\phi_1},
\nln
\smat_{12}\,\state{\phi_1\psi_2}\eq 
\frac{x^+_2-x^+_1}{x^-_2-x^+_1}\,\state{\psi_2\phi_1}
+\frac{x^+_2-x^-_2}{x^-_2-x^+_1}\,\frac{q_1}{q_2}\,\state{\phi_2\psi_1},
\nln
\smat_{12}\,\state{\psi_1\phi_2}
\eq
\frac{x^-_2-x^-_1}{x^-_2-x^+_1}\,\state{\phi_2\psi_1}
+\frac{x^+_1-x^-_1}{x^-_2-x^+_1}\,\frac{q_2}{q_1}\,\state{\psi_2\phi_1},
\nln
\smat_{12}\,\state{\psi_1\psi_2}\eq 
-\frac{x^-_2-x^+_1}{x^-_2-x^+_1}\,\state{\psi_2\psi_1}.
\>
This is precisely the S-matrix found in 
\cite{Beisert:2005fw}
for the $\alg{su}(1|2)$ sector of $\superN=4$ SYM. 
There, the functions $x^\pm(u)$ are defined intrinsically
through the equations
\[\label{eq:xpmdef}
\frac{x^+}{x^-}=\exp(ip),
\qquad
x^++\frac{\lambda}{x^+}-x^--\frac{\lambda}{x^-}=i.
\]
In particular, this means that the energy $e$ of an excitation,
which is related to its central charge via $c=1+\lambda e$,
is given by
\[\label{eq:disp}
e=\frac{i}{x^+}-\frac{i}{x^-}\,.
\]
This agrees with the dispersion relation used in \cite{Beisert:2005fw}.


\section{XXZ Spin Chain}

In conventional nearest-neighbour spin chains the representation 
of all excitations is the same. In particular, it does not
depend on the momentum and therefore one might set $c=1$ for all
excitations. In that case 
\[
x^\pm=a\pm \sfrac{i}{2}
\]
and the S-matrix \eqref{eq:su11s}
reduces to that of the standard $\alg{su}(2|1)$ spin chain 
with nearest-neighbour interactions (we can also set $q_1=q_2$).
\medskip

Inspired by the simplicity of \eqref{eq:su11s}
one might try to employ the notation using $x^\pm$ 
also to other well-known cases. 
Let us investigate the XXZ spin chain 
which is a deformation of the $\alg{su}(2)$ Heisenberg chain
with two spin orientations labelled by $\state{\fldZ}$ and 
$\state{\phi}$. The Hamiltonian reads
\[
\ham_{12}=
\quarter(r^{+1}+r^{-1})(\ident_{1}\ident_{2}-\sigma^z_1\sigma^z_2)
-\half(\sigma^x_1\sigma^x_2+\sigma^y_1\sigma^y_2)
\]
The anisotropy parameter is $\half (r^{+1}+r^{-1})$.
Alternatively, we shall investigate a similar spin chain 
where we replace the bosonic state $\state{\phi}$
by the fermionic $\state{\psi}$.
Using the above representation 
\eqref{eq:extrans} with $c=q=1$ and $b=0$ the Hamiltonian reads
\[
\ham_{12}=
\half(r^{+1}+r^{-1})(\ident_{1}\ident_{2}-\half\acomm{\gen{B}_1}{\gen{C}_2})-\comm{\gen{Q}_1}{\gen{S}_2}.
\]
The nearest-neighbour Hamiltonians
$\ham_{12}$ act on a pair of spins as%
\footnote{The interaction 
$\state{\fldZ\phi}\to +\state{\fldZ\phi}$,
$\state{\phi\fldZ}\to -\state{\phi\fldZ}$
is a boundary contribution and can be dropped.}
\[
\begin{array}[b]{rcl}
\ham_{12}\state{\fldZ\fldZ}\eq 0,
\\[3pt]
\ham_{12}\state{\fldZ\phi}\eq 
r^{-1}\state{\fldZ\phi}
-\state{\phi\fldZ},
\\[3pt]
\ham_{12}\state{\phi\fldZ}\eq 
r^{+1}\state{\phi\fldZ}
-\state{\fldZ\phi},
\\[3pt]
\ham_{12}\state{\phi\phi}\eq 
0,
\end{array}\qquad
\begin{array}[b]{rcl}
\ham_{12}\state{\fldZ\fldZ}\eq 0,
\\[3pt]
\ham_{12}\state{\fldZ\psi}\eq 
r^{-1}\state{\fldZ\psi}
-\state{\psi\fldZ},
\\[3pt]
\ham_{12}\state{\psi\fldZ}\eq 
r^{+1}\state{\psi\fldZ}
-\state{\fldZ\psi},
\\[3pt]
\ham_{12}\state{\psi\psi}\eq 
(r^{+1}+r^{-1})\state{\psi\psi}.
\end{array}
\]

It is straightforward to perform the coordinate space Bethe ansatz
for these systems. 
We can now define the $x^\pm$ as
\[\label{eq:xxzmom}
e^{ip}=\frac{x^+}{rx^-}\,,\qquad
r^{-1}x^+-r^{+1}x^-=\ihalf(r^{+1}+r^{-1}).
\]
The second equation permits us to solve both $x^\pm$ 
in terms of one spectral parameter $u$
\[\label{eq:xxzxpm}
x^\pm=r^{\pm 1}u\pm\ihalf.
\]
The dispersion relation is then similar to \eqref{eq:disp}
\[\label{eq:xxzdisp}
e=\half \lrbrk{r^{+1}+r^{-1}}\lrbrk{\frac{i}{x^+}-\frac{i}{x^-}}
\]
and the S-matrix (a factor) is simply
\[\label{eq:SmatXXZ}
\smat_{12}\,\state{\phi_1\phi_2}=\frac{x_2^+-x_1^-}{x_2^--x_1^+}\,\state{\phi_2\phi_1}\,,
\qquad
\smat_{12}\,\state{\psi_1\psi_2}=-\state{\psi_2\psi_1}\,.
\]
for bosons and fermions, respectively.
Consequently, the Bethe equations read
\[
\lrbrk{\frac{rx_k^-}{x_k^+}}^L \mathop{\prod_{j=1}^K}_{j\neq k} \frac{x_k^+-x_j^-}{x_k^--x_j^+}=1,\qquad
\lrbrk{\frac{rx_k^-}{x_k^+}}^L =1.
\]

Let us relate our parametrisation to the 
common one involving trigonometric functions,
cf.~\cite{Gaudin:1971aa,Alcaraz:1987uk,Sklyanin:1988yz} and the review \cite{Faddeev:1996iy}.
The variables $x^\pm$ are then related to a spectral parameter $\nu$ as follows
\[
x^\pm=\coth(\kappa)\sin\bigbrk{\kappa(\nu \pm \sfrac{i}{2})}\exp\bigbrk{-i\kappa(\nu \pm \sfrac{i}{2})},
\qquad 
r=\exp(\kappa).
\]
Furthermore, the parameter $r$ has been replaced by $\kappa$.
The propagation phase in the trigonometric form reads
\[
e^{ip}=\frac{x^+}{rx^-}=\frac{\sin\bigbrk{\kappa(\nu+\sfrac{i}{2})}}{\sin\bigbrk{\kappa(\nu-\sfrac{i}{2})}}
\]
and the scattering term becomes a function
of the difference of spectral parameters
\[
\frac{x^+_1-x^-_2}{x^-_1-x^+_2}=\frac{\sin\bigbrk{\kappa(\nu_1-\nu_2+i)}}{\sin\bigbrk{\kappa(\nu_1-\nu_2-i)}}\,.
\]
The dependence on $\nu_1-\nu_2$ 
is the benefit of the trigonometric parametrisation;
if this is not desired, one can employ the algebraic formulation
in \eqref{eq:SmatXXZ}, cf.~also
\cite{Frolov:2005ty} for a similar parametrisation.

\section{Quantum Algebra Deformed Spin Chains}

The algebraic parametrisation of XXZ-like spin chains
also generalises to symmetry algebras of higher rank. 
Here we consider deformations of chains with 
spins transforming the fundamental
representation of $\alg{su}(3)$ and $\alg{su}(2|1)$.
The generic model with these properties has
recently been investigated in \cite{Freyhult:2005ws}. 
The three states of a spin are 
$\state{\fldZ},\state{\phi},\state{\psi}$,
the first two are always bosonic and the statistics of the
third depends on the algebra.
The Hamiltonian acts on two different spins as follows \cite{Freyhult:2005ws}
\[\begin{array}[b]{rcl}
\ham_{12}\state{\fldZ\phi}\eq 
r^{-1}\state{\fldZ\phi}
-\state{\phi\fldZ},
\\[3pt]
\ham_{12}\state{\fldZ\psi}\eq 
r^{-1}\state{\fldZ\psi}
-\state{\psi\fldZ},
\\[3pt]
\ham_{12}\state{\phi\psi}\eq 
r^{-1}\state{\phi\psi}
-\state{\psi\phi},
\end{array}\qquad
\begin{array}[b]{rcl}
\ham_{12}\state{\phi\fldZ}\eq 
r^{+1}\state{\phi\fldZ}
-\state{\fldZ\phi},
\\[3pt]
\ham_{12}\state{\psi\fldZ}\eq 
r^{+1}\state{\psi\fldZ}
-\state{\fldZ\psi},
\\[3pt]
\ham_{12}\state{\psi\phi}\eq 
r^{+1}\state{\psi\phi}
-\state{\phi\psi}.
\end{array}
\]
For two equal spins we assume for $\alg{su}(3)$ 
\[
\ham_{12}\state{\fldZ\fldZ}=0,\qquad
\ham_{12}\state{\phi\phi}=0,\qquad
\ham_{12}\state{\psi\psi}=0
\]
and for $\alg{su}(2|1)$
\[
\ham_{12}\state{\fldZ\fldZ}=0,\qquad
\ham_{12}\state{\phi\phi}=
0,\qquad
\ham_{12}\state{\psi\psi}=
(r^{+1}+r^{-1})\state{\psi\psi}.
\]

In the coordinate space Bethe ansatz we can use the 
above definition of parameters $x^\pm$
\eqref{eq:xxzmom,eq:xxzxpm}
and obtain the same dispersion relation
\eqref{eq:xxzdisp}
for both excitations $\phi,\psi$ above
the vacuum of $\fldZ$'s, cf.~\cite{Freyhult:2005ws}.
Even more, the S-matrix also takes 
almost the above form \eqref{eq:su11s}.
The scattering of $\phi$ and $\psi$ merely
picks up factors of $r$
\<
\smat_{12}\,\state{\phi_1\psi_2}\eq 
r^{-1}\,\frac{x^+_2-x^+_1}{x^-_2-x^+_1}\,\state{\psi_2\phi_1}
+\frac{x^+_2-x^-_2}{x^-_2-x^+_1}\,\frac{q_1}{q_2}\,\state{\phi_2\psi_1},
\nln
\smat_{12}\,\state{\psi_1\phi_2}
\eq
r^{+1}\,\frac{x^-_2-x^-_1}{x^-_2-x^+_1}\,\state{\phi_2\psi_1}
+\frac{x^+_1-x^-_1}{x^-_2-x^+_1}\,\frac{q_2}{q_1}\,\state{\psi_2\phi_1}.
\>
The scattering of identical particles for $\alg{su}(3)$ is
\[
\smat_{12}\,\state{\phi_1\phi_2}=
\frac{x^+_2-x^-_1}{x^-_2-x^+_1}\,\state{\phi_2\phi_1},
\qquad
\smat_{12}\,\state{\psi_1\psi_2}=
\frac{x^+_2-x^-_1}{x^-_2-x^+_1}\,\state{\psi_2\psi_1},
\]
whereas for $\alg{su}(2|1)$ we get the same as in \eqref{eq:su11s}
\[
\smat_{12}\,\state{\phi_1\phi_2}=
\frac{x^+_2-x^-_1}{x^-_2-x^+_1}\,\state{\phi_2\phi_1},
\qquad
\smat_{12}\,\state{\psi_1\psi_2}=
-\frac{x^-_2-x^+_1}{x^-_2-x^+_1}\,\state{\psi_2\psi_1}.
\]
Note, however, that this 
S-matrix satisfies the Yang-Baxter equation 
only if the relation \eqref{eq:xxzmom} 
for $x^+$ and $x^-$ is fulfilled. 
This is in contrast with the S-matrix \eqref{eq:su11s}
which satisfies the YBE even if $x^+$ and $x^-$ are
independent.

The above S-matrix for deformed $\alg{su}(2|1)$ agrees with the 
one obtained in \cite{Bracken:1994hz,Delius:1994yn,Zhang:1998bn} 
when setting $\alpha=\beta=1$ and identifying 
\[
r=q,\qquad
x^\pm=\frac{i}{2}\,\frac{q^{+1}+q^{-1}}{q^{+1}-q^{-1}}\,\lrbrk{q^{\pm 1}x^{-1}-1}.
\]

The nested Bethe ansatz for this system leads to
a spectral parameter $y^\pm$ for the auxiliary Bethe roots
\[
y^\pm=r^{\pm1}(v\pm \ihalf).
\]
For the main Bethe equation we obtain
\[
\lrbrk{\frac{rx_k^-}{x_k^+}}^L \mathop{\prod_{j=1}^K}_{j\neq k} \frac{x_k^+-x_j^-}{x_k^--x_j^+}
\prod_{j=1}^{J} r^{+1}\frac{x_k^--v_j}{x_k^+ - v_j}=1.
\]
For $\alg{su}(3)$ and $\alg{su}(2|1)$, 
the auxiliary equations read, respectively
\[
\mathop{\prod_{j=1}^J}_{j\neq k} \frac{y_k^+-y_j^-}{y_k^--y_j^+}
\prod_{j=1}^{K} r^{+1}\frac{y_k^--u_j}{y_k^+ - u_j}=1,
\qquad
\prod_{j=1}^{K} r^{+1}\frac{y_k^--u_j}{y_k^+ - u_j}=1.
\]
%

\section{Conclusions}

In this note we have studied the S-matrix and asymptotic Bethe equations
for a particular long-range spin chain with $\alg{su}(2|1)$ symmetry 
which arises in a sector of planar $\superN=4$ super Yang-Mills theory. 
The model has the interesting feature that 
the representation of an excitation
depends on its momentum along the spin chain. 
This property is reflected by the S-matrix which takes
an interesting generic form using the spectral parameters $x^\pm$. 
We have also considered nearest-neighbour chains 
with quantum deformed $\alg{su}(3)$ and $\alg{su}(2|1)$ 
symmetries. Their S-matrices and Bethe ans\"atze are
usually formulated using trigonometric functions, 
but an alternative formulation using
$x^\pm$-parameters leads to very similar structures 
as for the above $\alg{su}(2|1)$ long-range chain.

We hope these results may improve the understanding 
of generic long-range spin chain models \cite{Beisert:2005wv}
and their relation to the well-known nearest-neighbour chains.
The results may also be useful for the study
of quantum strings on $AdS_5\times S^5$
by means of Bethe ans\"atze \cite{Arutyunov:2004vx}, 
spin chains \cite{Beisert:2004jw}
and supersymmetric subsectors \cite{Alday:2005jm,Arutyunov:2005hd}.
Finally, the findings can be generalized to 
the complete $\superN=4$ SYM spin chain model \cite{Beisert:2005aa}.

\ifarxiv\paragraph{Acknowledgements.}\else\section*{Acknowledgements}\fi
I thank Andreas Kl\"umper, Charlotte Kristjansen, Nicolai Re\-she\-tik\-hin,
Matthias Staudacher and Yao-Zhong Zhang for discussions and 
pointing out references. 
This work is supported in part by the U.S.~National Science
Foundation Grant No.~PHY02-43680. 
Any opinions, findings and conclusions or recommendations expressed in this
material are those of the author and do not necessarily reflect the
views of the National Science Foundation.

\ifarxiv\bibliographystyle{nbshort}\else\bibliographystyle{nbshorth}\fi
\bibliography{su11s}

\end{document}